\documentclass{aa}
\usepackage{natbib}
\usepackage{graphicx}
\usepackage[varg]{txfonts}
\usepackage{hyperref}
\hypersetup{citecolor=black, urlcolor=blue, linkcolor=black, colorlinks=true}
\newcommand{\sect}[1]{Sect.\,\ref{#1}}
\newcommand{\sects}[1]{Sects.\,\ref{#1}}
\newcommand{\fig}[1]{Fig.\,\ref{#1}}
\newcommand{\figs}[1]{Figs.\,\ref{#1}}

\newcommand{\eqn}[1]{Eq.\,(\ref{#1})}

\begin{document}

%
\title{Plasmoid-mediated reconnection in solar UV bursts} 

%
\authorrunning{H.~Peter et al.}

\author{H.~Peter\inst{1}, Y.-M. Huang\inst{2,3}, L. P. Chitta\inst{1}, and P. R. Young\inst{4,5}}

\institute{Max Planck Institute for Solar System Research, 
           37077 G\"ottingen, Germany, email: peter@mps.mpg.de
           \and
           Max-Planck-Princeton Center for Plasma Physics,
           Princeton, NJ 08540, USA
           \and
           Department of Astrophysical Sciences and 
           Princeton Plasma Physics Laboratory,
           Princeton University, Princeton, NJ 08540, USA
           \and
           NASA Goddard Space Flight Center,
           Solar Physics Laboratory, Greenbelt, MD 20771, USA
           \and
           Northumbria University, Newcastle Upon Tyne, NE1 8ST,
           UK
}

\date{Received 2 May 2019 / Accepted 1 July 2019}

\abstract%
%
{%
Ultraviolet bursts are transients in the solar atmosphere with an increased impulsive emission in the extreme UV lasting for one to several tens of minutes. They often show spectral profiles indicative of a bi-directional outflow in response to magnetic reconnection.
}
{%
To understand UV bursts, we study how motions of magnetic elements at the surface can drive the self-consistent formation of a current sheet resulting in plasmoid-mediated reconnection. In particular, we want to study the role of the height of the reconnection in the atmosphere.
}
{%
We conducted numerical experiments solving the 2D magnetohydrodynamic equations from the solar surface to the upper atmosphere. Motivated by observations, we drove a small magnetic patch embedded in a larger system of magnetic field of opposite polarity. This type of configuration creates an X-type neutral point in the initial potential field. The models are characterized by the (average) plasma-$\beta$ at the height of this X point. 
}
{%
The driving at the surface stretches the X-point into a thin current sheet, where plasmoids appear, accelerating the reconnection, and a bi-directional jet forms. This is consistent with what is expected for UV bursts or explosive events, and we provide a self-consistent model of the formation of the reconnection region in such events. The gravitational stratification gives a natural explanation for why explosive events are restricted to a temperature range around a few 0.1\,MK, and the presence of plasmoids in the reconnection process provides an understanding of the observed variability during the transient events on a  timescale of minutes.

}
{%
Our numerical experiments provide a comprehensive understanding of UV bursts and explosive events, in particular of how the atmospheric response changes if the reconnection happens at different plasma-$\beta$, that is, at different\ heights in the atmosphere. This analysis also gives new insight into how UV bursts might be related to the photospheric Ellerman bombs.
}
%
%

\maketitle

\section{Introduction\label{S:intro}}

The solar atmosphere is in a highly dynamic state. Since the early days of space-based spectroscopy at extreme ultraviolet (EUV) and X-ray wavelengths, observations have  hinted at small-scale dynamics that are either barely or not resolved. 
Using data from rocket flights, \cite{1983ApJ...272..329B} isolated what they originally called turbulent events and what were later termed explosive events.
They stood out as excessive broadening of spectral lines interpreted as nonresolved motions in response to heating of the plasma.
The broadened spectral features often appeared with additional components in the the red and blue wings of the line profile about 50\,km\,s$^{-1}$ to 100\,km\,s$^{-1}$ away from line center \cite[][]{1989SoPh..123...41D}.
The spatial location of bi-directional line profiles has  been connected to places of flux emergence, cancelation, and therefore reconnection \cite[][]{1991JGR....96.9399D}.
Given the limited amount of observing time of the experiments on rockets and a Space Shuttle flight, these explosive events have been mostly seen in the quiet Sun where a clear connection could be established to the bright network structures and the magnetic field \cite[][]{1991ApJ...370..775P}.
With the era of the Solar and Heliospheric Observatory \cite[SOHO;][]{1995SoPh..162.....F}, observations became much more abundant.
This allowed \cite{1997Natur.386..811I} to show that the misalignment of the line of sight with the axis of the reconnection outflow explains the asymmetry of the line profiles and to confirm the reconnection scenario.
This is consistent with the finding that explosive events occur mostly at locations where opposite polarities come into close contact and cancel \cite[][]{1998ApJ...497L.109C}.
For a review of the zoo of transient events in the UV, see \cite{2018SSRv..214..120Y}.

In active regions, in particular while magnetic flux is still emerging, energetic events are seen more frequently.
Forming in the photosphere, Ellerman bombs are visible through enhancements in the wings of H$\alpha$ \cite[][]{1917ApJ....46..298E}.
Observations give clear indications that these too are forming through reconnection  \cite[e.g.,][]{2002ApJ...575..506G}.
Recent observations in the UV reveal a new type of event that can be characterized by very strong enhancements in chromospheric and transition region lines (e.g., \ion{Mg}{II}, \ion{Si}{IV}).
Here spectral absorption features of chromospheric lines can be seen in the strongly broadened transition region line profiles \cite[][]{2014Sci...346C.315P}.
These UV bursts seem to originate in the chromosphere, thus from higher up (at lower densities) than Ellerman bombs.
While the plasma in UV bursts is heated sufficiently  to give rise to emission in, for example, \ion{Si}{IV}, the plasma temperature probably does not reach coronal temperature of the order of 1\,MK.

Explosive events have been seen abundantly in the quiet Sun in emission lines forming (under ionization equilibrium conditions) at around 0.1\,MK to 0.4\,MK, for example, in \ion{Si}{IV}, \ion{C}{IV,} or \ion{O}{VI}.
They are one of the classical transition-region phenomena.
They seem not to appear at higher temperatures (in the corona) in that they have no counterparts   in the upper transition region or low corona, for example in \ion{Ne}{VIII} forming at about 0.7\,MK or \ion{Mg}{X} at about 1\,MK \cite[][]{2002A&A...392..309T}.
This poses the question of why the mechanism that works in the transition region would not be efficient in the corona.

After it had been established observationally that explosive events are caused by reconnection, 2D magnetohydrodynamics (MHD) models were constructed with the aim to recreate the observed bi-directional reconnection outflows.
Most of the models so far have used a Harris-type current sheet \cite[][]{Harris:1962} as an initial setup.
For example, \cite{1999SoPh..185..127I} used a magnetic field that changes as $\tanh$ across the current sheet.
To initiate reconnection, these early models used an anomalously enhanced resistivity at a single location along the current sheet.
The resulting reconnection outflow speeds and temperatures are consistent with observations \cite[][]{1999SoPh..185..127I} in that they provide an explanation for the enhanced emission in the line wings often showing up as separate spectral components \cite[][]{2001A&A...370..298R,2001A&A...375..228R,2001A&A...380..719R}.
When also considering gravitational stratification, \cite{2002A&A...383..697R} were able to find asymmetric line profiles \cite[as seen in observations;][]{2004A&A...427.1065T}, but they could still not recover the increase of the intensity in the line core that is associated with most (but not all) transient events.

This shortcoming of the early explosive event models has been worked out clearly by \cite{2015ApJ...813...86I} and they suggested that the presence of plasmoids in the current sheet provides a solution.
Direct observational evidence for plasmoids has been found in the solar atmosphere, at least in currents sheets forming following a coronal mass ejection \cite[][]{2003ApJ...594.1068K,2005ApJ...622.1251L}, in flares \cite[][]{2004ApJ...605L..77A,2012ApJ...745L...6T}, in the corona \cite[e.g.,][]{2016NatPh..12..847L}, and recently also in the chromosphere \cite[][]{2017ApJ...851L...6R}. The key importance of the formation of plasmoids for reconnection was established early on \cite[][]{1986PhFl...29.1520B}.
For large Lundquist numbers, thin current sheets will undergo a super-Alfv\'enic tearing instability \cite[e.g.,][]{2007PhPl...14j0703L,2009PhPl...16k2102B} which results in a rich variety of possible reconnection dynamics \cite[e.g.,][]{2013PhPl...20e5702H}.
The importance of plasmoid-mediated reconnection for the solar chromosphere and corona became evident quickly \cite[e.g.,][]{2012ApJ...758...20N,2015ApJ...799...79N,2012ApJ...760..109L}. Using models of plasmoid-mediated reconnection, \cite{2015ApJ...813...86I} showed that this indeed can explain the strong enhancement of the line cores along with the increased emission in the line wings.
This is because the plasmoid instability provides not only faster reconnection (outflows), but also more plasma at slow speeds in the forming plasmoids.
One main reason why the early explosive event models did not show the plasmoid instability is a lack of spatial resolution.

The 2D reconnection modeling is now also applied to UV bursts and Ellerman bombs by placing the reconnection region in places of different plasma-$\beta$, where $\beta$ is the ratio of magnetic to thermal energy density, or magnetic to gas pressure.
If $\beta$ is lower, that is, higher in the atmosphere, the plasma can be heated to higher temperatures, and in the low chromosphere gravity can hinder the formation of plasmoids \cite[][]{2016ApJ...832..195N}.
However, the situation changes if one accounts for the presence of neutrals in the chromosphere and the reconnection in the low chromosphere can result in peak temperatures of about 30\,kK \cite[][]{2018ApJ...852...95N,2018ApJ...868..144N}.
During reconnection, the 2D models can produce turbulent small-scale structures in plasmoids that cover a range of temperatures from $10^4$\,K to $10^5$\,K, that is, covering the temperatures expected in Ellerman bombs and UV bursts.
Still,  whether or not the observational signatures expected from those models are consistent with observations remains unknown.
Observations show that there is a relation between Ellerman bombs and UV bursts \cite[][]{2015ApJ...812...11V}.
In a significant fraction of cases (10\% to 20\%) Ellerman bombs and UV bursts appear at the same time and location \cite[][]{2016ApJ...824...96T}.

The first 3D MHD models to reproduce the radiative signatures of Ellerman bombs were presented in \cite{2017A&A...601A.122D} and \cite{2017ApJ...839...22H}, and the latter model was also able to reproduce UV burst emission from \ion{Si}{iv}, but unrelated to the Ellerman bomb at a different location.
In a very recent 3D model, \cite{2019A&A...626A..33H} find a UV burst flashing at the same time and location as the Ellerman bomb.
While these models provide important insight into the evolution and driving of these events, the resolution is (naturally) not comparable to the 2D models and thus they barely resolve the process of plasmoid-mediated reconnection, if at all.

From the above discussion we can summarize some of the major shortcomings and open questions for understanding explosive events and UV bursts.
(1) Current 2D models assume the existence of a Harris-type current sheet as the initial setup of the model.
How is this current sheet produced in the first place?
(2) How is the response of the atmosphere changing when the site of the reconnection is located at sites of different plasma-$\beta$?  
(3) Why are explosive events seen only at transition region temperatures and not above approximately 0.6\,MK in the (low) corona?
(4) How are explosive events, UV bursts, and Ellerman bombs connected in terms of magnetic configurations and models?
Are they based on the same mechanism, despite the fact that they seem to appear in regions where plasma-$\beta$ is above and below unity?

Our study addresses questions (1) to (3). Concerning question (4), we can only provide an educated guess. We address these questions using a 2D model with sufficient resolution to resolve the plasmoids that form in the current sheet.
The main improvements over existing 2D models are as follows: we use a magnetic field setup that is motivated by solar observations; the simulation is driven by a horizontal motion at the bottom boundary, that is, the solar surface; and this driving is inspired by observations.
Our setup is guided by the work of \cite{2017A&A...605A..49C} who found a small magnetic patch of one magnetic polarity moving between two large regions of opposite polarity.
A magnetic field extrapolation of the observation revealed an X-type neutral point above the small magnetic patch and the spectroscopic EUV observations showed the continued presence of a UV burst above that patch.
We capture this magnetic setup in our model, drive the small patch into the larger region of magnetic field, and find that the X-point stretches into a current sheet.
We therefore create the thin current sheet in a self-consistent way. With the plasmoid instability setting in, we get significant energy conversion that we can relate to the observed UV burst. Running numerical experiments with different plasma-$\beta$ we investigate how the atmosphere reacts in these cases and check the peak temperatures during the reconnection events.

\section{Magnetohydrodynamic model setup\label{S:model}}

The idealized 2D setup of the model is inspired by observations of UV bursts.
Therefore, we first give a general figurative explanation of the setup of the magnetic field and how it is driven from the surface (\sect{S:setup}).
Subsequently, we present the MHD equations solved here together with the initial and boundary conditions (\sects{S:model.eqs} to \ref{S:boundary}) and discuss the range of parameters for our numerical experiments (\sect{S:experiments}).

\subsection{General setup\label{S:setup}}

The magnetic setup is inspired  by a UV burst observation near a site of flux cancelation by \cite{2017A&A...605A..49C}.
In their study they related the emission from the UV burst to the magnetic field extrapolated into the upper atmosphere.
They found the UV burst to be located at an X-type magnetic null point.
Essentially, this X-point was located some 500\,km above a (minor) parasitic polarity that moved in a region of opposite polarity \cite[see Fig.\,8 of][]{2017A&A...605A..49C}.

\begin{figure}
\centerline{\includegraphics[width=75mm]{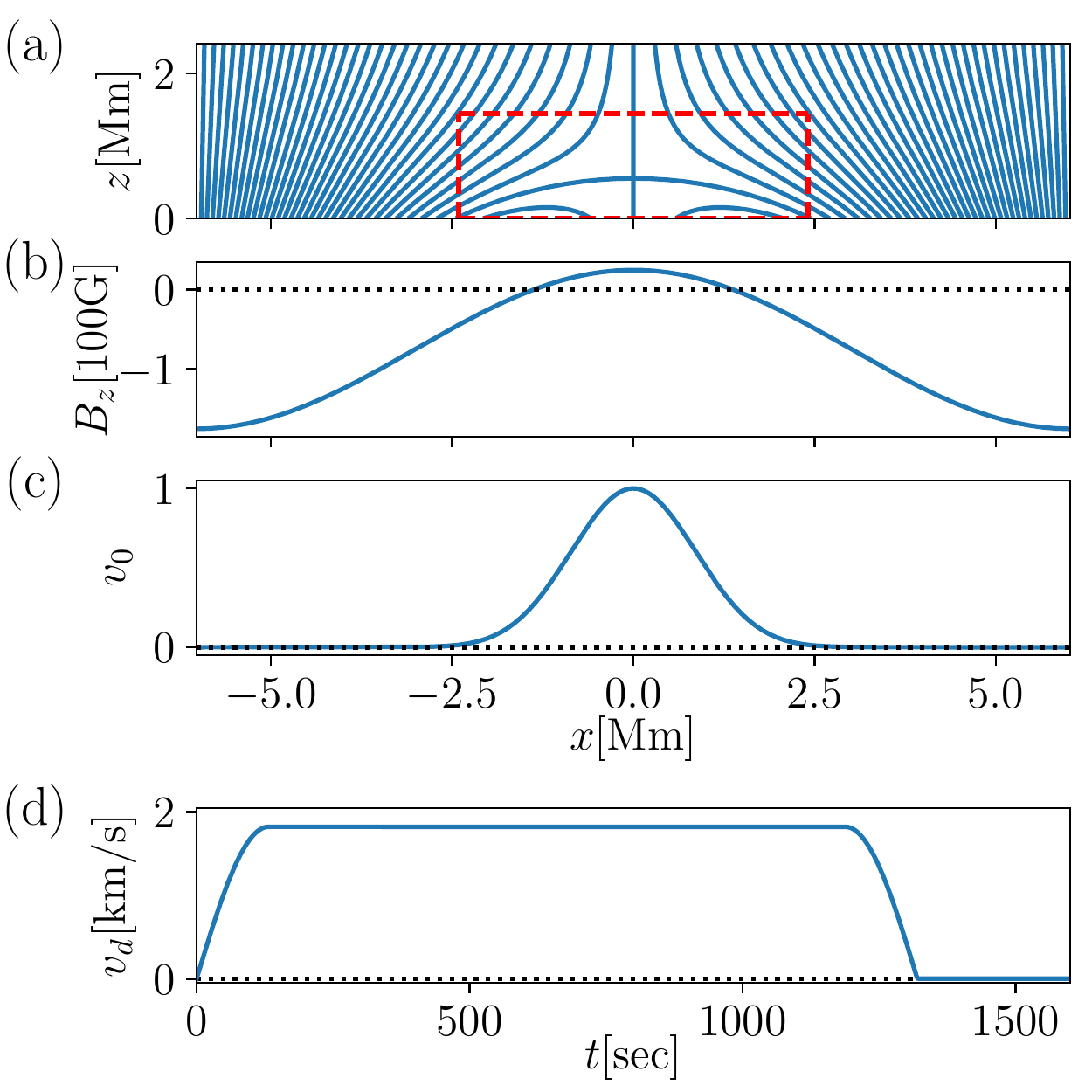}}
\caption{Model setup.
In panel (a) we show the initial condition of the magnetic field through the field lines in the 2D computational domain following \eqn{E:B.initial}. 
Panel (b) displays the vertical component of the magnetic field at the bottom boundary at $z{=}0$.
The magnetic field is driven in the positive $x$-direction with a Gaussian-type velocity profile $v_0(x)$ as defined in \eqn{E:v.space} and depicted in panel (c).
The peak value of the driving velocity $v_d(t)$, given through \eqn{E:v.time}, is shown in panel (d).
While panel (a) shows the full computational domain, the red dashed rectangle indicates the region of interest displayed in \fig{F:evolution}. 
See \sect{S:model} for details. 
\label{F:setup}}
\end{figure}

In our 2D model, the initial magnetic field at the lower boundary reflects this setup (cf.\ \fig{F:setup}a).
The whole computational domain stretches over 12\,Mm in the horizontal ($x$) and 2.4\,Mm in the vertical ($z$) direction.
To mimic the magnetic setup from \cite{2017A&A...605A..49C}, we choose a magnetic field as defined later in \eqn{E:B.initial} where at the bottom boundary the main (negative) polarity covers the region outside $|x|\gtrsim1.2$\,Mm and the small patch of parasitic (positive) polarity is in the center of the box inside $|x|\lesssim1.2$\,Mm.
This is illustrated by the field lines and the vertical component of the magnetic field at the bottom boundary shown in \fig{F:setup}a,b.
Clearly, this magnetic field configuration has a magnetic null point in the form of an X-point above the parasitic polarity at a height of about 550\,km above the bottom boundary as can be seen from \fig{F:setup}a and detailed in \eqn{E:null}.
This would correspond to a location in the chromosphere on the real Sun.

In their observations, \cite{2017A&A...605A..49C} reported that the small parasitic polarity moved from one main polarity into the other main polarity.
In our model we mimic this by applying a horizontal motion in the $+x$ direction.
The normalized spatial profile $v_0(x)$ of the velocity driver is shown in \fig{F:setup}c and detailed later in \eqn{E:v.space}.
We restrict the driving to the middle part of the computational domain for the simple reason that on the real Sun the opposite polarity would be pushed only so far into the main polarity before it gets completely canceled, while the main polarity would remain (roughly at the same location).
Furthermore, we restrict the peak value of the driving velocity, $v_d(t)$, in time for some 20~minutes, as illustrated in \fig{F:setup}d and defined in \eqn{E:v.time}.
Starting with the stable initial potential magnetic field setup, we slowly ramp up the velocity for about two~minutes at the bottom boundary and similarly ramp down to reduce the velocity at the end of the driving period to zero.
Driving with a maximum speed of just below 2\,km\,s$^{-1}$ (which is compatible with the observations) will move the parasitic polarity by about 2\,Mm during the period of driving.

As expected, when running the numerical experiments, the driving of the magnetic field at the bottom boundary moves the patch of opposite polarity in the $+x$ direction.
Consequently, the null point above the opposite polarity also moves in the same direction \cite[as found also in the observation by][]{2017A&A...605A..49C}.
The X-point then stretches into a current sheet and plasmoid-mediated reconnection begins (cf. \fig{F:evolution}).

\subsection{Model equations\label{S:model.eqs}}

The governing equations for our simulations are the 2D resistive MHD equations in a vertical $x$-$z$ plane.
Here we solve for the magnetic potential $\psi$ instead of the magnetic field itself, meaning that $\mathbf B=\nabla\psi\times\hat{y}$, where $\hat{y}$ is the unit vector in the $y$ direction.
The induction equation then reads
\begin{equation}\label{E:induction}
\partial_{t}\psi+\mathbf{v}\cdot\nabla\psi=\eta\nabla^{2}\psi,
\end{equation}
where $\mathbf{v}$ denotes the velocity vector and $\eta$ is the (constant) magnetic resistivity.
In the dimensionless units we chose $\eta$ to be $10^{-5}$, which in real units corresponds to $4.4\times10^5$\,m$^2$\,s$^{-1}$.
This value is about a factor of 300 larger than the value according to classical transport theory \cite[e.g.,][]{Boyd+Sanderson:2003} for temperatures at about $10^4$\,K, that is, at the typical background temperature of our models (cf.\ \sect{S:init}).
We chose this anomalously high value to ensure that the current sheets are dissipated at the grid scale (the vertical grid spacing $\delta{z}$ is just above 1\,km; see \sect{S:init}).
For the velocities of about $U\approx1$\,km\,s$^{-1}$ driving the system (cf.\ \sect{S:boundary}), $U\,\delta{z}$ would be of the same order of magnitude as $\eta$, and consequently the magnetic Reynolds number on the grid scale would be about unity. This ensures efficient dissipation at the grid scale and avoids numerical dissipation.

The continuity and momentum equations are given by
\begin{equation}\label{E:mass}
\partial_{t}\rho+\nabla\cdot\left(\rho\mathbf{v}\right)=0,
\end{equation}
\begin{equation}\label{E:momentum}
\partial_{t}(\rho\mathbf{v})+\nabla\cdot\left(\rho\mathbf{vv}\right)=-\nabla p-{\rho}g\hat{z}-\nabla\psi\nabla^{2}\psi+\nabla\cdot(\rho\nu\mbox{\boldmath$E$\unboldmath}),
\end{equation}
where we assume the pressure of an ideal gas, $p=2n{k_{\rm{B}}}T$, with number density $n{=}\rho/m_{\rm{p}}$, Boltzmann's constant $k_{\rm{B}}$,
and temperature $T$. The mass density is denoted by $\rho$ and the proton mass by $m_{\rm{p}}$.
The gravitational acceleration $g=274$\,m\,s$^{-2}$ operates in the negative $z$ direction ($\hat{z}$ is the unit vector in the $+z$ direction). The Lorentz force is given through $\nabla\psi\nabla^{2}\psi$.
The effects of viscosity are included through the strain rate tensor
$\mbox{\boldmath$E$\unboldmath}=(\nabla\mathbf{v}+\nabla\mathbf{v}^{T})/2$
and the kinematic viscosity $\nu$.
To choose
the value for the viscosity $\nu$, we use the same reasoning  as above for the resistivity $\eta$. Consequently they have the same values, $\nu{=}\eta$.

Finally, the energy balance is described by
\begin{equation}\label{E:energy}
\partial_{t}p+\nabla\cdot(p\mathbf v)=(\gamma-1)\left(-p\nabla\cdot\mathbf{v}-\nabla\cdot\mathbf{q}+H_{\eta}+H_{\nu}\right).
\end{equation}
Heat conduction parallel to the magnetic field is through the heat flux $\mathbf{q}=-\kappa_{\parallel}\mathbf{bb}\cdot\nabla T$, with $\mathbf{b}$ being the unit vector along the magnetic field.
The coefficient of the thermal conductivity is given by $\kappa_{\parallel}=c_{\rm{v}}\rho\chi$, with the specific heat at constant volume $c_{\rm{v}}$, and the thermal diffusivity $\chi$.
 For simplicity, we assume a constant diffusivity of $\chi=4.4\times10^9$\,m$^2$\,s$^{-1}$. %
According to classical transport theory, $\kappa_{\parallel}$ should depend
on temperature as ${\propto}T^{5/2}$. Considering this temperature dependence, the typical values of $\chi$ at the top of the chromosphere, the transition region and the corona would be of the order of $10^5$\,m$^2$\,s$^{-1}$, $10^8$\,m$^2$\,s$^{-1}$, and $10^{11}$\,m$^2$\,s$^{-1}$, respectively.%
\footnote{Assuming temperatures of $10^4$\,K, $10^5$\,K, and $10^6$\,K in an isobaric atmosphere with a typical coronal density of $10^{9}$\,cm$^{-3}$ ($\hat{=}10^{-12}$\,kg\,m$^{-3}$) at $10^6$\,K and using $\kappa_{\parallel}=10^{-11} (T[{\rm{K}}])^{5/2}$\,W\,m$^{-1}$\,K$^{-1}$ \cite[e.g.,][]{Priest:1982}.}
Thus the value of $\chi$ we use in our model is higher than expected for the top of the chromosphere or the transition region, that is, the regions where the UV bursts or explosive events we intend to model are located.
The higher efficiency of heat conduction in our model implies that the temperatures we find during the reconnection events will be a lower limit.

Energy is added to the plasma through Ohmic dissipation, $H_{\eta}=\eta j^{2}$ (with currents $j{=}|\nabla{\times}B|)$, and viscous dissipation $H_{\nu}=n\nu\,\nabla\mathbf{v}\,{:}\,\mbox{\boldmath$E$\unboldmath}$.
The ratio of specific heats is denoted by $\gamma{=}5/3$.
We neglected optically thin radiative losses, because the timescale of the events is short compared to the (coronal) radiative cooling times.

The code that is used to solve the above equations under the boundary conditions discussed below has been described in \cite{1993PhFlB...5.3712G}. More details on the code as well as on testing and validation are given there.

\subsection{Setup and initial conditions\label{S:init}}

The simulation stretches in the horizontal direction over $x\in[-L_{x},L_{x}]$ and in the vertical direction over $z\in[0,L_{z}]$.
We chose $L_x=6$\,Mm and $L_z=2.4$\,Mm motivated by the observations of \cite{2017A&A...605A..49C}; see \sect{S:setup}.
The computational domain is covered by 2400 grid points horizontally and 2000 grid points vertically.
Consequently the (equidistant) grid is spaced by $5\,{\rm{km}}\times1.2\,{\rm{km}}$.

The analytical form of the initial magnetic field, $\mathbf B_0=\nabla\psi_0\times\hat{y}$, is defined through
\begin{equation}\label{E:B.initial}
\psi_{0}(x,z;t{=}0)=\left[f_b\frac{L_{x}}{\pi}\sin\left(\frac{\pi x}{L_{x}}\right)\exp\left(\frac{-\pi z}{L_{x}}\right)-x \right] \overline{B}.
\end{equation}
This represents a potential magnetic field, that is, one can easily show that $\nabla\times\mathbf{B}_0=0$, and of course, $\nabla\cdot\mathbf{B}_0=0$.
Most importantly, this magnetic field structure as depicted in \fig{F:setup}a roughly matches the field geometry motivated by the UV burst observations of \cite{2017A&A...605A..49C}.
In particular it has an X-type null point at 
\begin{equation}\label{E:null}
x_{\rm{null}}=0 \quad , \quad z_{\rm{null}}=\left(L_{x}/\pi\right)~\ln f_b.
\end{equation}
Considering the horizontal extent of our box, $2L_x=12$\,Mm, we choose the free parameter $f_b=4/3$ so that the height of the null point is $z_{\rm{null}}\approx550$\,km.
This then closely matches the height of the null point derived by \cite{2017A&A...605A..49C}.
In the form of \eqn{E:B.initial}, the average vertical magnetic field in the model region is $\overline{B}$.
Here we choose $\overline{B}=75$\,G, which is typical for a plage or enhanced network area, where UV bursts are seen frequently.

We choose the temperature at the beginning of the simulation to be constant.
This is motivated by our goal to simulate a UV burst that happens at the base of the corona.
Embedded in an comparatively cool environment the plasma should be heated to transition region temperatures, as expected for a UV burst.
Consequently our constant initial temperature is set to $T_0=10^4$\,K.
The initial state is at rest, that is, all velocities are zero.

For the density we choose an initial state with a vertical (barometric) stratification in the atmosphere with the initially constant temperature,
\begin{equation}\label{E:init.density}
\rho\,(x,z;t{=}0) =\rho_{0}\exp\left(-z/H_g\right),
\end{equation}
where $H_g=(2k_{\rm{B}}T_0)/(m_{\rm{p}}\,g)$ is the barometric scale height.
In our setup this is about $H_g\approx600$\,km.
The density at the bottom boundary, $\rho_0$, is a free parameter and is chosen differently in the numerical experiments conducted here (see.\ \sect{S:experiments}).

\subsection{Boundary conditions\label{S:boundary}}

To drive the system, we apply a spatially and temporally variable velocity profile at the bottom boundary. This velocity is always in the positive $x$ direction.
The spatially variable part is defined as
\begin{equation}\label{E:v.space}
v_0(x) = \left(\frac{1+\cos\left(\pi x/L_{x}\right)}{2}\right)^{\!\!10}.
\end{equation}
This closely resembles a Gaussian profile with a full width at half maximum of 2\,Mm for our choice of $L_x=6$\,Mm (cf. \fig{F:setup}c).
In contrast to a Gaussian, this form in \eqn{E:v.space} is strictly periodic and therefore satisfies our horizontal boundary conditions.
We keep driving for a timescale of $t_d$. For a smooth transition we ramp the velocity up and down over the timescale $t_r$,
\begin{equation}\label{E:v.time}  
v_d(t)=\hat{v} \times
\left\{
\begin{array}{l@{~~~~~~}r@{~}c@{~}l}
\sin\left(0.5\pi t/t_{r}\right)        & 0 \le&t&\le t_{r}             \\
1                                      & t_{r} \le &t&\le t_{d}-t_{r}   \\
\sin\left(0.5\pi(t_{d}-t)/t_{r}\right) & t_{d}-t_{r} \le &t&\le t_{d}   \\
0                                      & & t&\ge t_d
\end{array}
\right.
.\end{equation}
Loosely guided by the observational study of \cite{2017A&A...605A..49C}, we choose the driving time to be $t_d=1320$\,s, or about 20 minutes, and apply a ramping time of one tenth of that, $t_r=132$\,s, to ensure a smooth transition.
The peak driving velocity is $\hat{v}=1.8$\,km\,s$^{-1}$, and therefore during the major part of the driving the velocity $v_{\rm{d}}$ is slightly faster than in \cite{2017A&A...605A..49C}. This makes up for the slightly shorter period of driving in our numerical experiments.
The time profile of the driving as defined in \eqn{E:v.time} is depicted in \fig{F:setup}d.
Combining the spatial and temporal variation in \eqn{E:v.space} and \eqn{E:v.time} yields the final horizontal velocity imposed at the bottom boundary,
\begin{equation}\label{E.v}
v_x\,(x,z{=}0;t) = v_d(t) ~\, v_0(x).
\end{equation}

The horizontal velocity at the top boundary and the vertical velocities at the top and bottom boundaries are set to zero.
In a real system, the plasma outflow from the reconnection region should be free to escape along the spine through the top boundary.
In the numerical model, we set the flow to zero at the top boundary for convenience and place a frictional layer near the top boundary to absorb the flow.
This mimics the outflow boundary condition and ensures that the plasma outflow does not bounce back from the top boundary and feedback to the reconnection site.  
The magnetic field at the bottom and top boundaries is line-tied to the flow field that drives the system.
The temperatures at the top and the bottom boundaries are fixed at their initial values. The density above and below the top and bottom boundaries, respectively, is extrapolated by a local gravitational stratification, ${\partial}p/{\partial}z=-\rho\,g$ (to fill the ghost cells).
The boundary conditions in the horizontal direction are periodic.

\subsection{Range of numerical experiments\label{S:experiments}}

In our study we also want to investigate the effects that plasma-$\beta$, that is,\ the ratio of gas pressure to magnetic pressure, has on the UV burst resulting from the driving. 
To characterize $\beta$ for the different numerical experiments, we use the average $\langle\beta\rangle$ in the dome region in the initial condition.
This dome region we define as a horizontal section in $x\in[-2.4,+2.4]$\,Mm at the height where the X-type null point as defined in \eqn{E:null} is located.
In our model setup the average field strength in the dome region is $\sqrt{\langle{B}^2\rangle}\approx 52$\,G (we first average $B^2$, because we are interested in the average magnetic pressure to calculate the average $\beta$).

To change $\beta$ in our idealized setup we could change the average magnetic field $\overline{B}$ (and thus $\sqrt{\langle{B}^2\rangle}$), the density at the bottom boundary, $\rho_0$, or a combination of both.
Here we opt to change $\rho_0$.
In our models we choose $\rho_0$ so that the density in the initial condition at the X-type null point has values from $4.9\times10^{-10}\,{\rm{kg}}\,{\rm{m}}^{-3}$ to $2.4\times10^{-8}\,{\rm{kg}}\,{\rm{m}}^{-3}$.
These are typical chromospheric densities \cite[cf.\ e.g.,][]{1981ApJS...45..635V}.
Essentially, with the choice of $\rho_0$ we select whether the X-type neutral point is in the low, middle, or upper chromosphere.

This combination of initial density and magnetic field in the dome region together with the initial temperature of $10^4$\,K gives a range of $\langle\beta\rangle$ from $0.015$ to $0.735$ in the dome region (even though we also ran more models with lower and higher $\beta$).
The respective $\langle\beta\rangle$ values will be given with the figures and the discussion of the results.
These averages of $\beta$ in the dome region around the X-type null point apply only for the initial condition.
When the numerical experiments evolve, the density, temperature, and magnetic field changes self-consistently and the location of the X-point moves (and eventually stretches into a current sheet).
Still, the initial value of $\langle\beta\rangle$ is a good measure to characterize the (average) plasma-$\beta$ in and around the reconnection region.

In our numerical experiments we change the density in order to change plasma-$\beta$ while we keep the magnetic field fixed.
What can we expect if we keep the density fixed and change the magnetic field to run models for the same (average) plasma-$\beta$?
For a case with the same $\beta$ but higher magnetic field, the magnetic energy density (per volume) would be increased by the same factor as the density.
This is because the energy density is ${\propto}B^2$ and $\beta{\propto}B^2/\rho$.
Furthermore, we can expect the energy dissipation (or conversion) to increase with the magnetic energy density.
Consequently, the heating per particle (i.e., heat input divided by density) will be the same in two models with the same $\beta$, even if the magnetic field and the density are higher.
According to the energy equation, the (change of) temperature goes with the heating per particle.
Therefore, we can expect to have models with different $\beta $ but the same peak temperatures in the reconnection region, even if the magnetic field is stronger.
However, this does not hold if  radiative cooling and effects of partial ionization are also included.
Increasing the temperature would then become more difficult if the X-point were in a region at higher density (and higher magnetic field so that $\beta$ is the same).
This has been shown in the multi-fluid models of \cite{2018ApJ...868..144N}.

\section{Results\label{S:results}}

\subsection{General evolution of the magnetic structure and formation of plasmoids\label{S:res.general}}

\begin{figure}[t]
\centerline{\includegraphics[width=75mm]{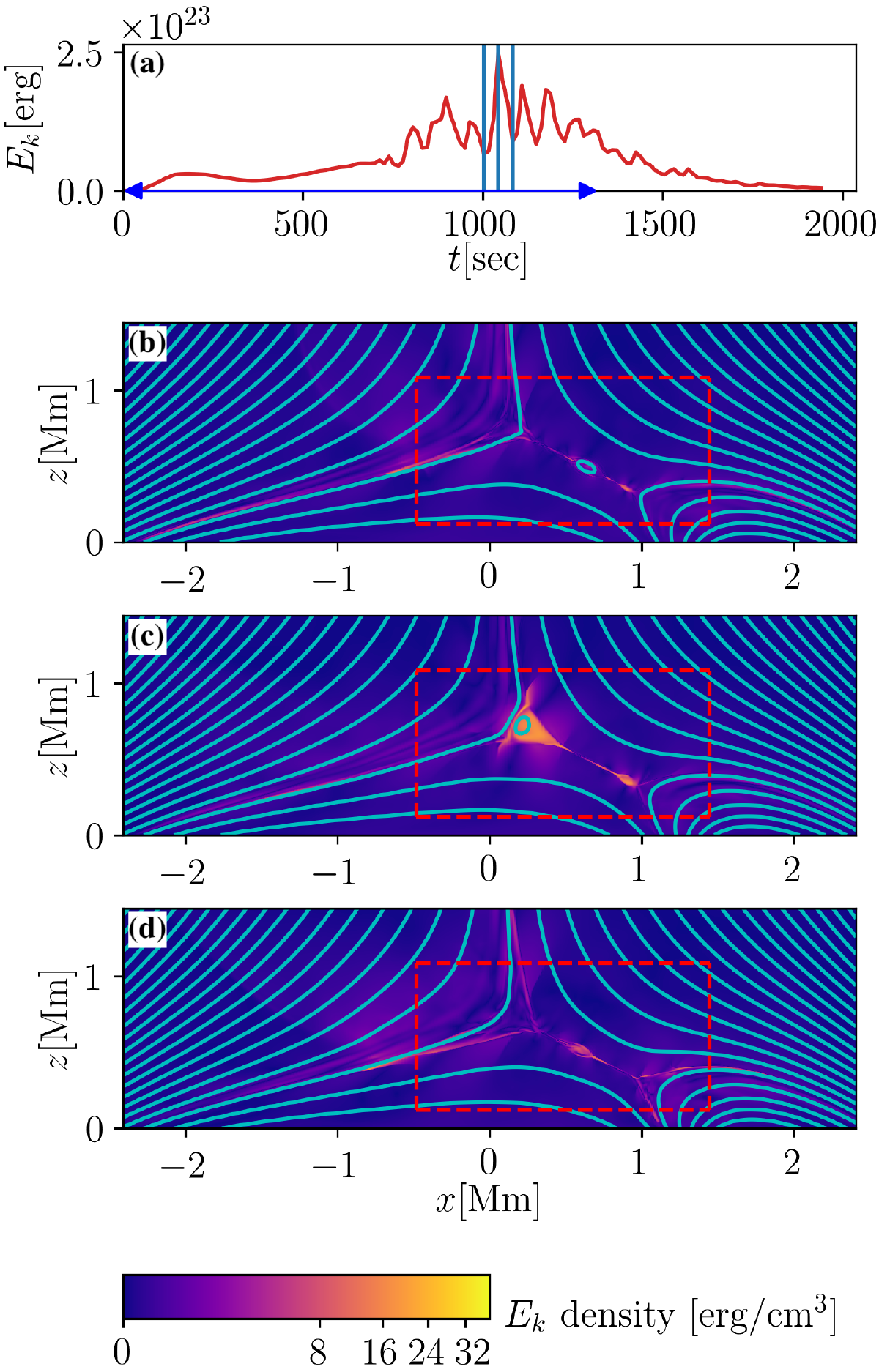}}
\caption{Overall evolution of  magnetic field and kinetic energy during reconnection for $\beta{=}0.147$.
Panel (a) shows the temporal evolution of the kinetic energy integrated in
a rectangle around the reconnection region (dashed lines in lower panels).
Panels (b) to (d) show snapshots of the kinetic energy density along with
magnetic field lines in part of the computational domain (cf. dashed rectangle
in \fig{F:setup}a).
These are taken sequentially at the times  indicated in the top panel by
the vertical blue lines.
The current sheet forms roughly along the diagonal of the rectangle. The
compact near-circular enhancements of the kinetic energy indicate the location
of the plasmoids that form.
The blue double-arrow in panel (a) indicates the time of the driving at the bottom boundary.
The plasmoids and the overall evolution are best seen in the animation attached
to this figure (also available at \url{http://www2.mps.mpg.de/data/outgoing/peter/papers/2019-bursts/f2.mp4});
see \sect{S:res.general}. 
\label{F:evolution}
}
\end{figure}

To describe the overall evolution of the system we first concentrate on one of the numerical experiments, and choose the case where $\beta{\approx}0.15$ at the height of the null point in the initial condition.
To illustrate the evolution of the system, we show in \fig{F:evolution} the kinetic energy together with the changing magnetic connectivity.
A movie attached to the figure shows the full temporal evolution. The lower panels (b) to (d) show part of the computational domain.
On the background with the kinetic energy density we plot the changing magnetic field lines.
The top panel (a) displays the kinetic energy integrated in a rectangle around the reconnection region as outlined by the dashed red lines in panels (b) to (d).

When driving the small opposite magnetic polarity in the positive $x$-direction (right), the magnetic field gets distorted and the X-type neutral point is stretched into a current sheet.
The current sheet is visible here through the  kinetic energy that is enhanced due to the reconnection outflow that roughly follows the diagonal of the dashed rectangles in \fig{F:evolution} (we discuss the currents later in \sect{S:res.beta} and \fig{F:current.sheets}).
We see a clear increase of the kinetic energy around the reconnection region due to the reconnection outflow associated with the motion of plasmoids (\fig{F:evolution}a).
After an initial phase, this enhancement of the kinetic energy essentially lasts as long as we drive the system (i.e., up to time $t{\approx}1300$\,s; cf. \fig{F:setup}d).
Details of the temporal evolution are presented in \sect{S:res.evolution}.

\begin{figure}[t]
\centerline{\includegraphics[width=75mm]{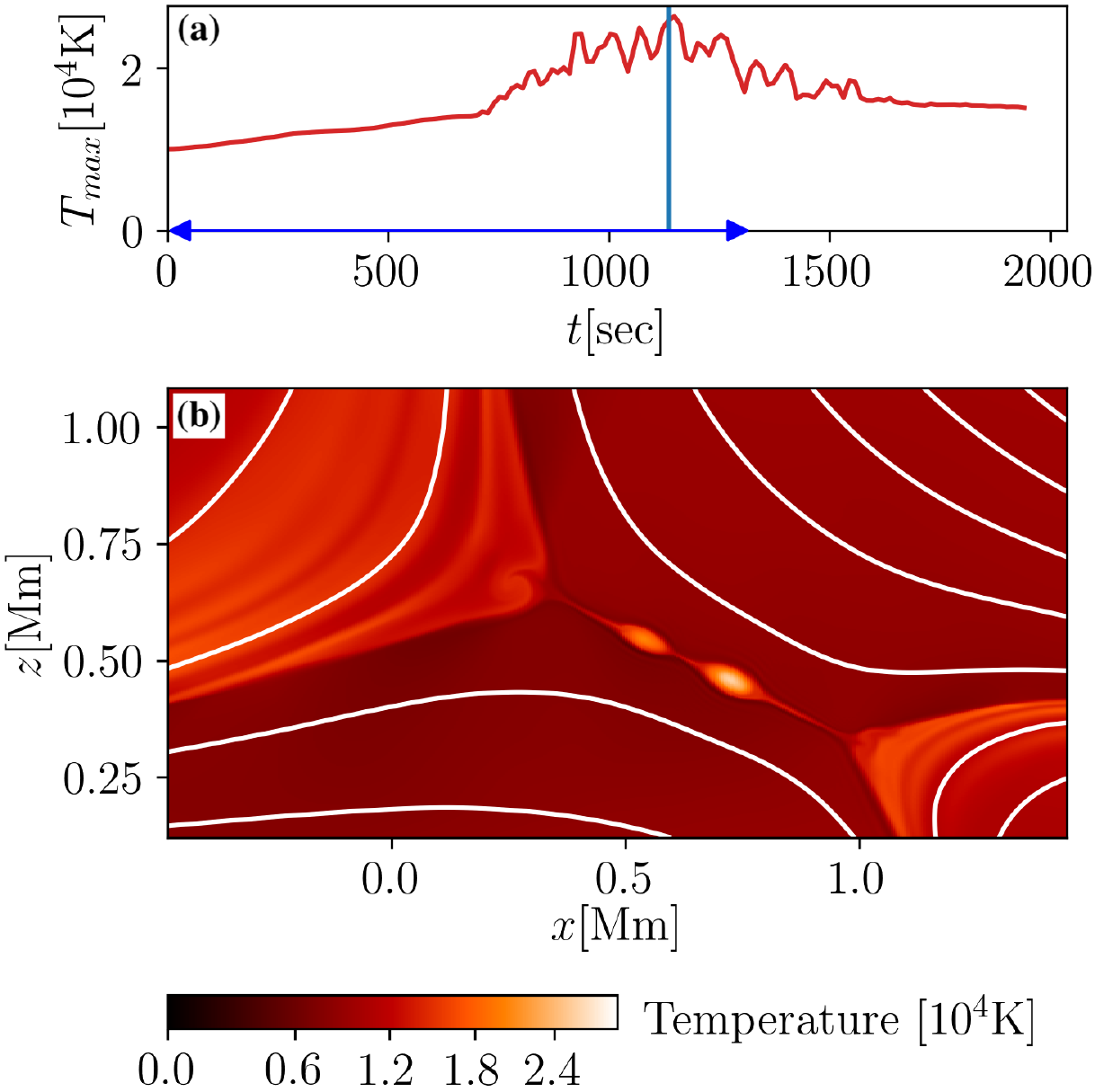}}
\caption{Zoom into reconnection region showing plasmoids and temperature for $\beta{=}0.147$.
This is similar to \fig{F:evolution}, but we now show the temporal evolution
of the maximum temperature around the reconnection region in panel (a) and a
snapshot of the  temperature in panel (b) that is taken at the time indicated by the blue vertical line in panel (a).
At this snapshot there are two (resolved) plasmoids present in the reconnection region clearly seen as near-circular regions of enhanced temperature in the middle of panel (b).
The field of view corresponds to the rectangle indicated in \fig{F:evolution}b-d.
The blue double-arrow indicates the time of the driving at the bottom boundary; see \sect{S:res.general}.
\label{F:zoom}
}
\end{figure}

The plasmoids form continuously and are regions of enhanced temperature and density.
This is underlined by \fig{F:zoom} where we show a zoom into the reconnection region.
In the snapshot we show here, two plasmoids are clearly visible as compact, near-circular enhancements of the temperature (see \sect{S:disc.beta} for a discussion of the size of plasmoids and resolution).
Also, the thin current sheet is visible as a thread of enhanced temperature running diagonally through the two plasmoids.
The plasmoids are hotter for a couple of reasons.
Ohmic heating is stronger in the current sheets, that is, in regions between plasmoids rather than in the plasmoids themselves.
The heated plasma is ejected into the plasmoids and trapped there.
The plasmoids then retain the heat because they are essentially thermally insulated.
This is because they are magnetic islands and thermal conduction is parallel to the field lines.
Furthermore, the plasma is heated by adiabatic compression due to the magnetic tension force in plasmoids.
This latter effect also causes the higher density within the plasmoids.
Consequently, the peak temperature (\fig{F:zoom}a) as well as the kinetic energy (cf.\ \fig{F:evolution}) are closely connected to the presence of plasmoids: whenever plasmoids are present there will be (local) maxima in temperature and in kinetic energy.
We discuss the resulting temporal substructure in \sect{S:disc.beta} and compare it to observations.

In the outflow regions (top left and bottom right regions in  \fig{F:zoom}b), that is, in the region that is fed by the reconnection outflow, we see a clear enhancement of the temperature as compared to the inflow regions.
 The case for the kinetic energy shown in \fig{F:evolution} is the same (but less clear).
As expected, the reconnection energizes  a region much larger than just the immediate surrounding of the current sheet.

\subsection{Temporal evolution of the reconnection and plasma-$\beta$\label{S:res.evolution}}

\begin{figure*}[t]
\parbox{183mm}{
\includegraphics[width=61mm]{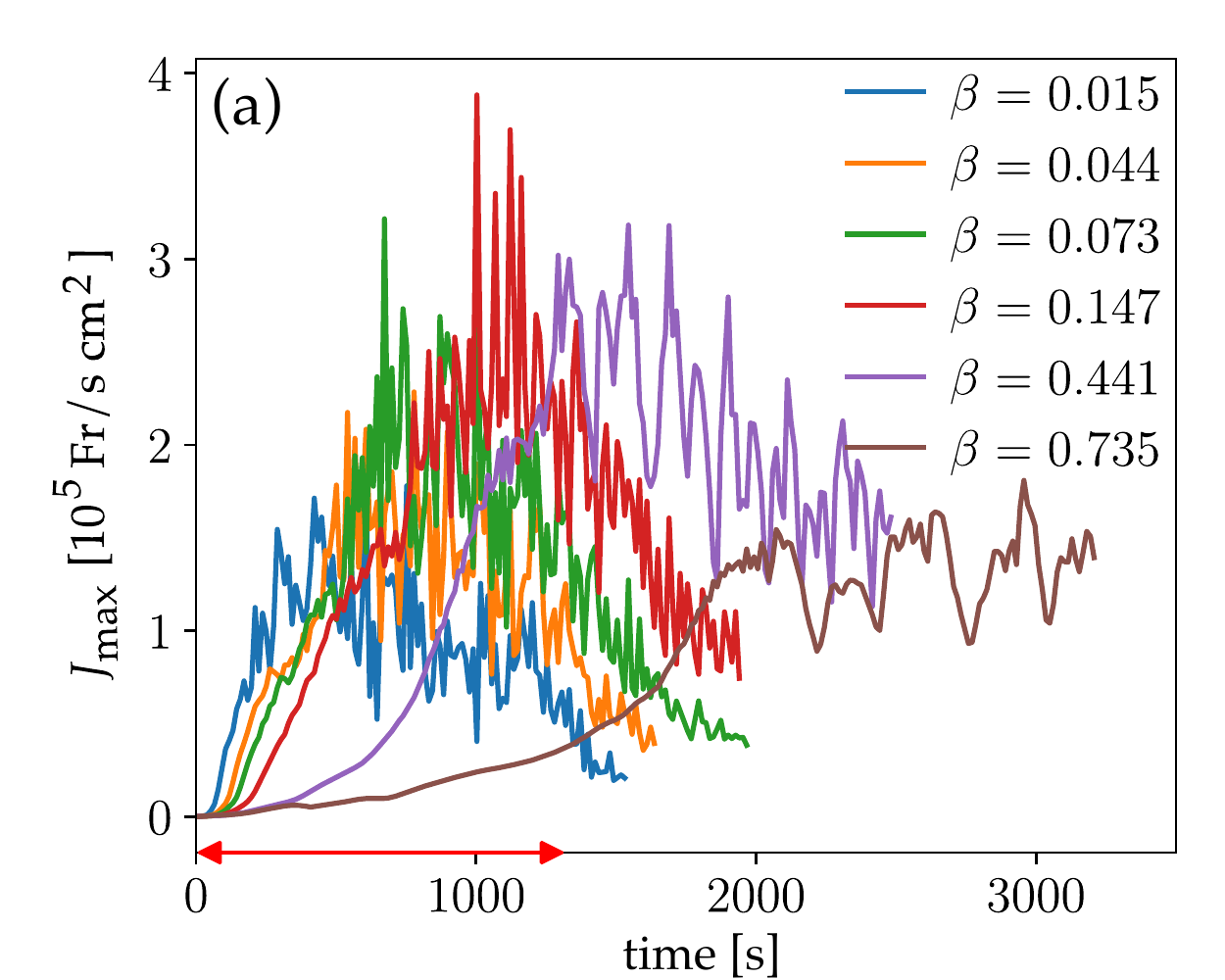}
\includegraphics[width=61mm]{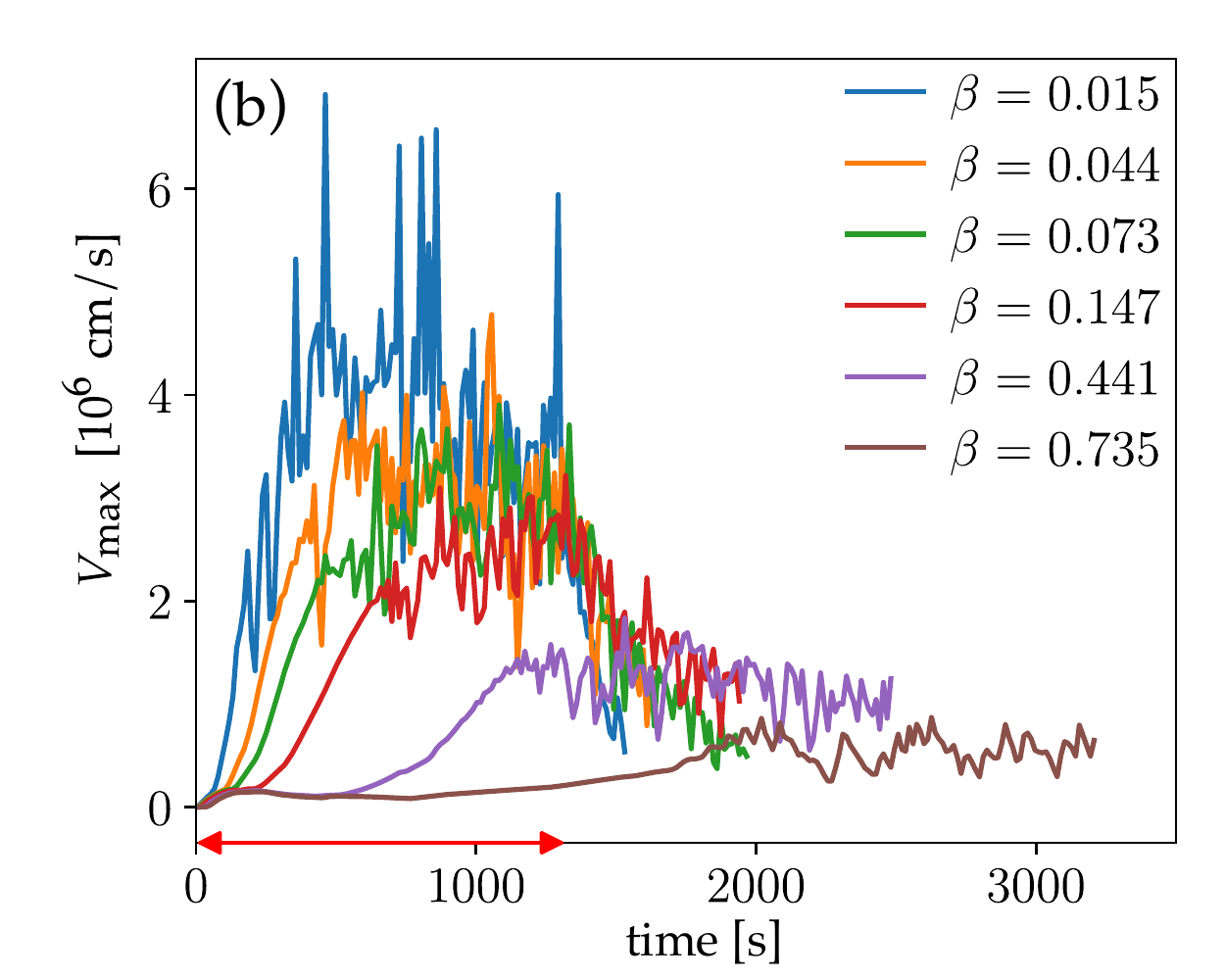}
\includegraphics[width=61mm]{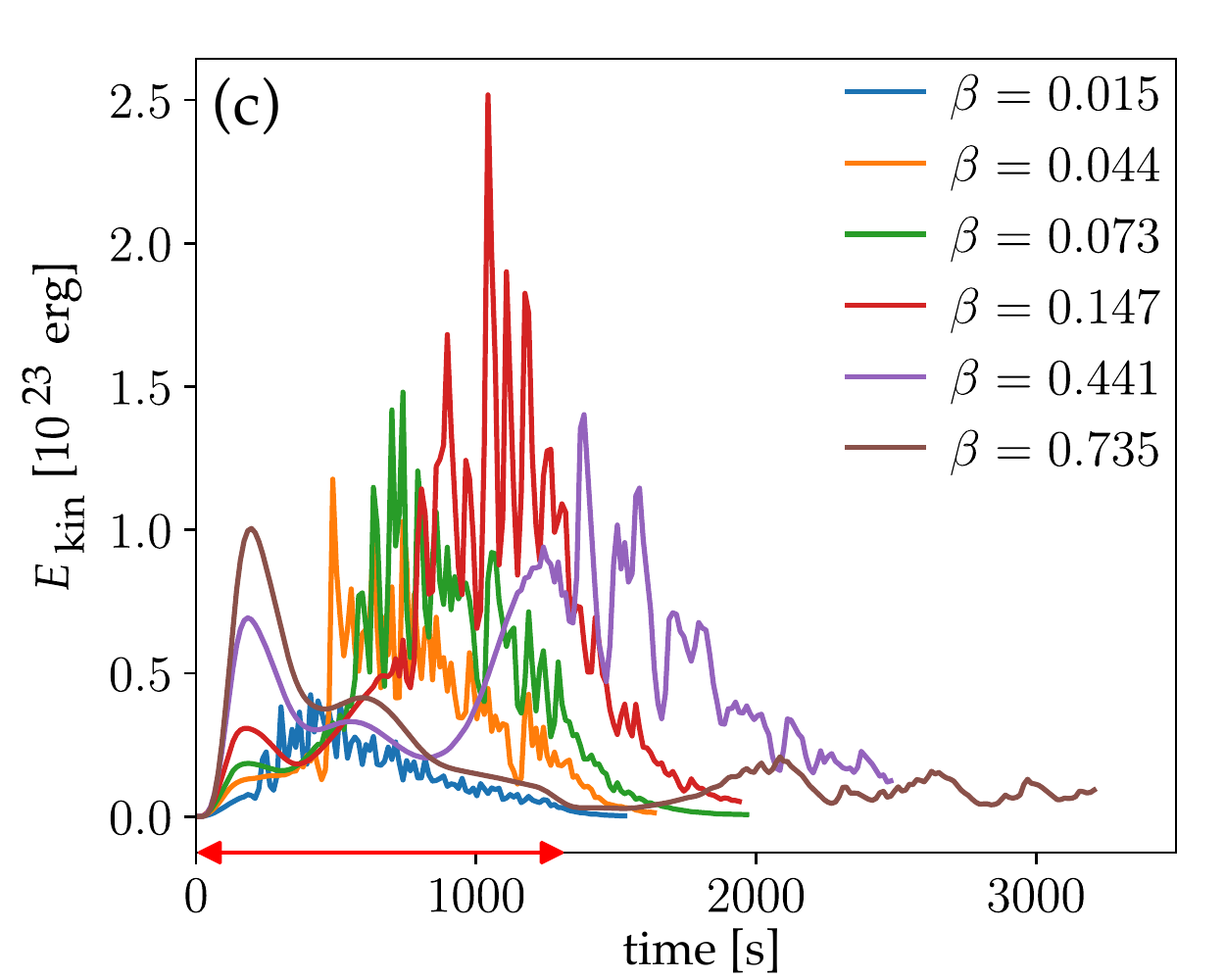}
}
\caption{Evolution of  reconnection for different plasma-$\beta$.
Here we show the temporal variation of the maximum current, $J_{\rm{max}}$ (a), and the maximum velocity, $V_{\rm{max}}$ (b), in the vicinity of the reconnection region as defined by the dashed rectangles in \fig{F:evolution}b-d.
Panel (c) shows the total kinetic energy, $E_{\rm{kin}}$, integrated over the same rectangle.
The differently colored curves show the evolution for the models with different plasma-$\beta$ in the reconnection region as outline in the legend.
The red double-arrows indicate the time of the driving at the bottom boundary. See \sect{S:res.evolution}. 
\label{F:evolution.beta}}
\end{figure*}

The different experiments we conduct are distinguished by the value of the average plasma-$\beta$ at the height of the reconnection region (at the initial condition; cf.\ \sect{S:experiments}).
To characterize how the reconnection evolves in the different cases, we consider a sub-region of size 2\,Mm $\times$ 1\,Mm that fully encloses the current sheet, the plasmoids, and the reconnection outflows (indicated by the dashed boxes in \fig{F:evolution}b-d). In this region we track changes of three characteristic quantities as a function of time, namely the maximum current, $J_{\rm{max}}$, the maximum velocity, $V_{\rm{max}}$, and the integrated kinetic energy, $E_{\rm{kin}}$.

The driving at the bottom boundary is the same for all the cases studied here (only the density stratification changes; see \sect{S:experiments}).
Therefore the difference in for example\ the energy deposition in the reconnection processes in the different cases is solely due to the different evolution of the magnetic field within the computational domain.
The experiments with values of plasma-$\beta$ from ${\approx}0.75$ to ${\approx}0.01$ range from (almost) plasma-dominated to magnetic-field-dominated in the vicinity of the reconnection region.
This then determines how efficiently reconnection can operate.

The overall evolution of the maximum current, maximum velocity, and kinetic energy for all the cases is similar (\fig{F:evolution.beta}).
These quantities increase, reach a peak while the driving at the footpoints is active, and then decline; they also show a substructure in the temporal evolution that looks almost like noise, but is real and is related to the presence of plasmoids.
This is discussed further when relating our models to observations in \sect{S:disc.beta}.
The high-$\beta$ cases are different in these quantities ($J_{\rm{max}}$, $V_{\rm{max}}$, $E_{\rm{kin}}$) in that they increase only later or do not reach a peak at all while driving.

\subsubsection{Current sheet and plasmoids}\label{S:curr.sheet.plasmoids}

For the maximum current, $J_{\rm{max}}$, we see a clear ordering of the timing of its sharp rise with plasma-$\beta$ (\fig{F:evolution.beta}a).
This is mainly due to the different Alfv\'en crossing times from the surface to the X-type neutral point, that will be longer for the larger $\beta$ values.
For lower $\beta$ values the Alfv\'en speed is higher and consequently the disturbance induced by the surface motions will reach the neutral point earlier and stretch it into a current sheet much sooner and quicker.

\begin{figure*}[t]
\includegraphics[width=184mm]{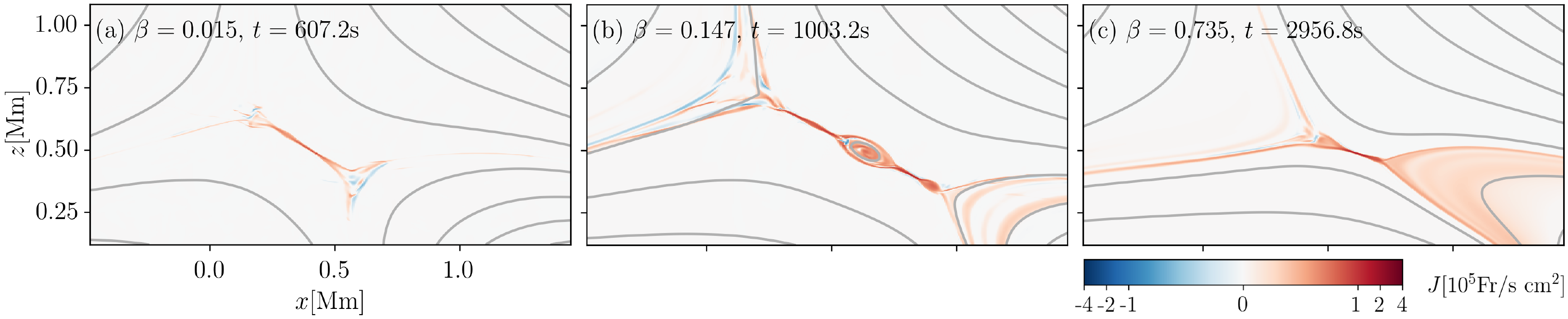}
\caption{Plasmoid-mediated reconnection under different plasma-$\beta$ conditions.
The panels show snapshots of the current density, $J$, for three different models with different values of plasma-$\beta$ according to values given in the plots.
The snapshots for low and moderate $\beta$ in panels (a) and (b) are taken around the time when the overall evolution of $J_{\rm{max}}$ reaches a maximum.
The snapshot for high $\beta$ in panel (c) represents the situation when the $J_{\rm{max}}$ reached stably high values (cf.\ \fig{F:evolution.beta}a).
See \sect{S:res.beta}. 
\label{F:current.sheets}}
\end{figure*}

For small $\beta$, below about 0.1, the current sheet forms essentially at the same time as the driving begins.
As a result, the current sheet will be shorter as compared to the experiments with higher $\beta$.
This is evident from a comparison of the snapshots of the currents in panels (a) and (b) of \fig{F:current.sheets} that show the cases for $\beta$ at about 0.015 and 0.15, respectively.

In the cases of high $\beta$, above about 0.4, the current sheet forms later not only because of the longer Alfv\'en crossing time, but also because the inertia of the plasma around the X-type neutral point can keep the magnetic field from forming a current sheet, at least for a while.
The resulting current sheets will be shorter at these higher values of $\beta$ (cf.\ \fig{F:current.sheets}c), because the stressing of the magnetic field has to work against the inertia of the plasma.
This also causes  longer-lasting significant magnetic gradients, that is, high currents lasting for longer (see case $\beta{=}0.735$ in \fig{F:evolution.beta}a).

In all models we find that plasmoids develop irrespective of the value for plasma-$\beta$ at the (initial) location of the X-type neutral point or the current sheet.
They can be seen in all quantities; in the current density (\fig{F:current.sheets}) of course, but also in the kinetic energy (\fig{F:evolution}) and in the temperature (\fig{F:zoom}a).
As expected, plasmoids will move away from the center of the current sheet and finally collide with the ambient medium and thermalize (see the animation attached to \fig{F:evolution} to see the motion of the plasmoids).

\subsubsection{Maximum velocity}

The reconnection outflow will be driven by the Lorentz force in the current sheet, with higher currents implying stronger forcing.
In the case of low $\beta$ the density in the vicinity of the current sheets will be smaller, and thus the Lorentz force can accelerate the gas to higher speeds.
Thus the resulting maximum velocity, $V_{\rm{max}}$, shows a comparable time variation as the maximum currents at the same $\beta$, but the highest $V_{\rm{max}}$ in each $\beta$ case drops with increasing $\beta$.
This is the simple consequence of the Lorentz force being applied to plasma at lower density.

\subsubsection{Kinetic energy\label{S:res.kinetic}}

The time variation of the kinetic energy, $E_{\rm{kin}}$, integrated in the vicinity of the reconnection region shows a two-part evolution: an increase at the beginning of the driving and then an evolution consistent with the maximum current and velocity.
In the initial phase of the driving, the horizontal motion at the bottom boundary simply carries the whole system in the positive $x$-direction and thus causes the initial bump of $E_{\rm{kin}}$ at $t{\approx}200$\,s (\fig{F:evolution.beta}c).
The speed of this motion is slow, of the order of less than 1 km\,s$^{-1}$ and therefore does not show up in the maximum velocity (\fig{F:evolution.beta}b), but because $E_{\rm{kin}}$ is integrated in a rather large volume, the bump is visible.
This small peak during the first 200\,s in $E_{\rm{kin}}$ is lower if $\beta$ is lower (and hence density is also lower). This simply reflects the scaling of the kinetic energy with the density.
While not being clear in the plot in \fig{F:evolution.beta}c, closer inspection shows that this early bump (at much smaller amplitude) is also present for the very small $\beta$ cases.
With the system finding some type of driving equilibrium, this increase of the kinetic energy vanishes again.
This initial bump of $E_{\rm{kin}}$, most prominent at high $\beta$, is an artifact of our driving and we do not discuss it further.

The most important part of the evolution of the kinetic energy, 
$E_{\rm{kin}}$, in the reconnection region is the main part that is closely related to the maximum velocity, $V_{\rm{max}}$.
The envelope of $V_{\rm{max}}$ reaches its highest values for small $\beta$ (\fig{F:evolution.beta}b).
In contrast, $E_{\rm{kin}}$ reaches
only small values for very low $\beta$ (\fig{F:evolution.beta}c). This is again because of the lower density for the low-$\beta$ cases.
Thus, with increasing $\beta$ also the maximum  $E_{\rm{kin}}$ increases, %
but only up to values of $\beta$ of around 0.1.
For higher values of $\beta$, the inertia of the plasma can work strong enough against the magnetic driving, so that the outflow velocities drops faster with $\beta$ than the density increases associated with higher $\beta$.
Consequently, for values of $\beta$ significantly above 0.1 we find only a small increase of $E_{\rm{kin}}$ during the reconnection event.

In addition to this evolution of the envelope of $E_{\rm{kin}}$ (and also $J_{\rm{max}}$ and $V_{\rm{max}}$), we find a substructure in the temporal variation with a timescale of about one minute (\fig{F:evolution.beta}) that looks almost like noise.
This substructure is clearly associated with the presence of plasmoids that form in the current sheet naturally through the plasmoid instability.
This can be seen upon close inspection of \fig{F:evolution}.
The snapshots in panels (b) and (d) are taken when there are no (or small) plasmoids in the current sheet.
At these times $E_{\rm{kin}}$ has local minima (\fig{F:evolution}a).
When there is a plasmoid (in the snapshot in \fig{F:evolution}c), there is a local maximum of $E_{\rm{kin}}$ in \fig{F:evolution}a.
This relation between plasmoids and local peaks of the kinetic energy is very clear when following the animation that is associated with \fig{F:evolution}. %
Likewise, there is a tight relation between the presence of plasmoids and $J_{\rm{max}}$ and $V_{\rm{max}}$ as well as the maximum temperature in the reconnection region, $T_{\rm{max}}$.
The timescale of the order of one minute  for this substructure is related to the time a plasmoid moves along the current sheet, bounded by the Alfv\'en speed. This is discussed further in \sect{S:disc.beta}.

\begin{figure*}
\sidecaption
\parbox[b]{12cm}{
\includegraphics[width=60mm]{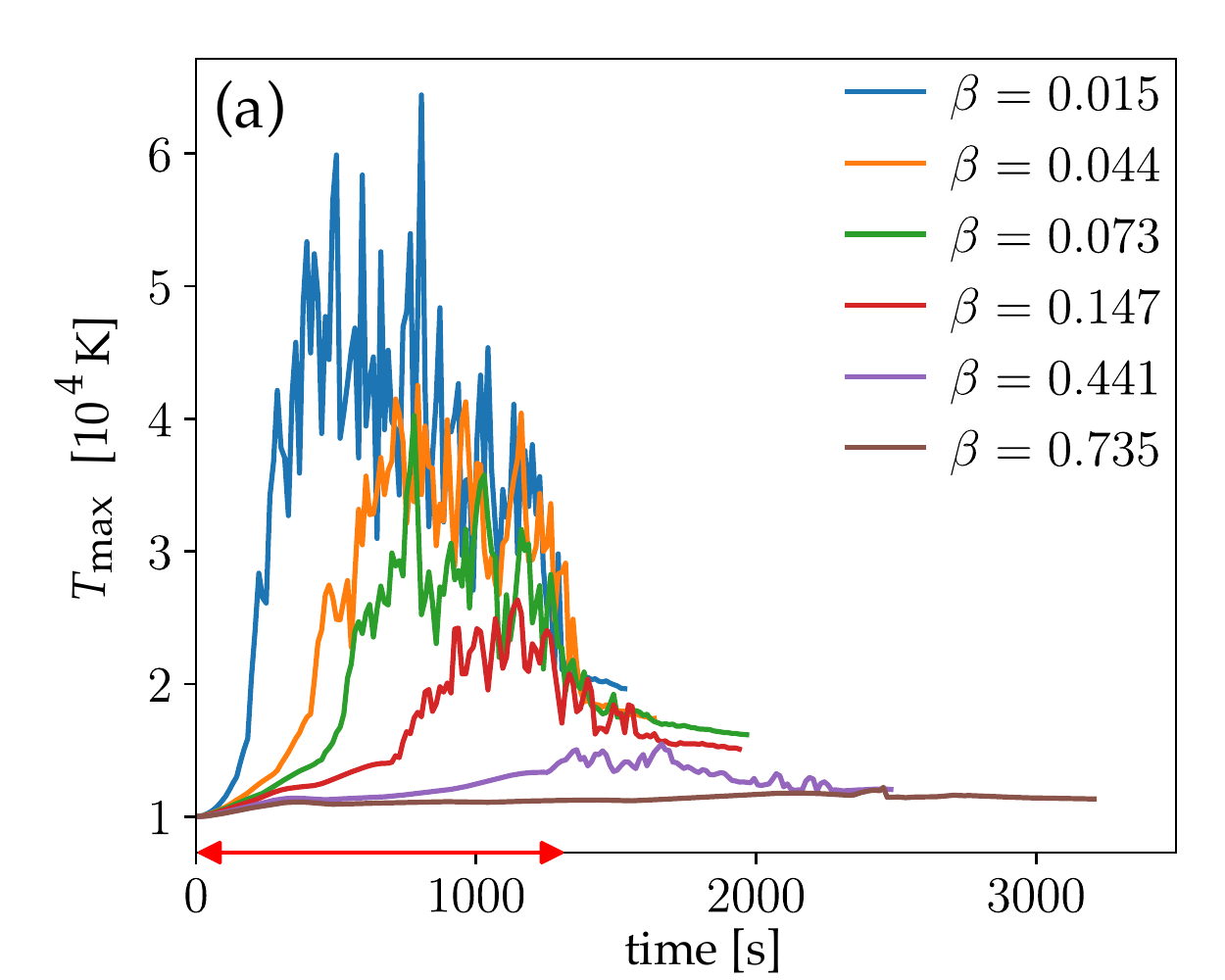}
\includegraphics[width=60mm]{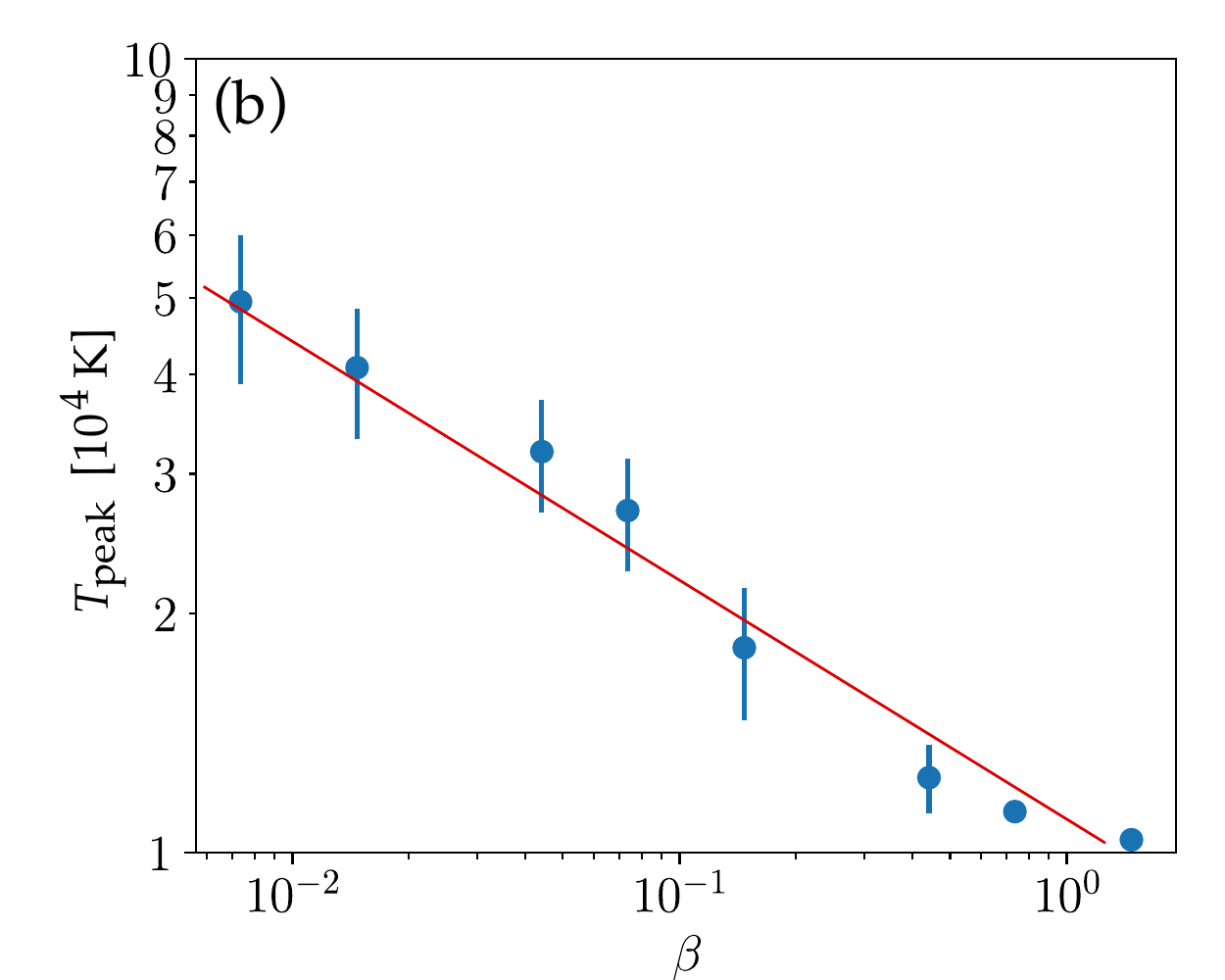}
}
\caption{Temperature and  plasma-$\beta$.
Panel (a) shows the evolution of the maximum temperature $T_{\rm{max}}$ in the vicinity of the reconnection region.
The differently colored lines display the models with different plasma-$\beta$ according to the legend.
This is similar to the plots in \fig{F:evolution}.
Panel (b) shows the peak temperature, $T_{\rm{peak}}$, (with the standard deviation of $T_{\rm{max}}$ shown as bars) for each of the runs as a function of plasma-$\beta$.
For illustrative purposes the red line in panel (b) indicates a power law $T_{\rm{peak}}{\propto}\beta^{\,-0.3}$.
See \sect{S:res.beta} for a definition of  $T_{\rm{peak}}$ and a more general discussion. 
\label{F:T.plasma.b}}
\end{figure*}

\subsection{Reconnection, plasma-$\beta$, and temperature\label{S:res.beta}}

We now turn to the peak temperatures during the reconnection event for the runs with different plasma-$\beta$.
In \fig{F:T.plasma.b}a we show the maximum temperature in the reconnection region as a function of time.
Here for each time we evaluated the maximum value of the temperature in the vicinity of the current sheet (i.e., in the rectangles shown in \fig{F:evolution}b-d, and the full region in \fig{F:zoom}b).
We see a clear trend that higher maximum temperatures are reached for lower plasma-$\beta$.

To quantify this, we estimate the peak temperature $T_{\rm{peak}}$ during the evolution of the maximum temperature in \fig{F:T.plasma.b}a for each run.
 The maximum temperature in the reconnection region, $T_{\rm{max}}$, varies rapidly because of the presence of plasmoids (\sect{S:curr.sheet.plasmoids}). Therefore we employ the following procedure to avoid spurious results due to individual peaks.  
We evaluate $T_{\rm{max}}$ during the time period after it first becomes larger than half of the overall maximum value, until it finally drops below half of the overall maximum value and never increases back again.
We define $T_{\rm{peak}}$ as the average of $T_{\rm{max}}$ over this period.
These values of $T_{\rm{peak}}$ are shown in \fig{F:T.plasma.b}b for the models with plasma-$\beta$ ranging from 0.007 to above unity.

We find that the peak temperature $T_{\rm{peak}}$ and plasma-$\beta$ are connected (roughly) by a power law,  $T_{\rm{peak}}\propto\beta^{\,-0.3}$ (cf.\ \fig{F:T.plasma.b}b).
For high plasma-$\beta$ we find no significant enhancement of the temperature, simply because the density is too high, meaning that the deposited energy can lead only to a modest increase in temperature.
The smaller plasma-$\beta$ , the higher the temperature enhancement can be, because the added energy is now distributed over many fewer particles.
In our models we can find the temperature rising to about 50\,000\,K.

\section{Discussion\label{S:discussion}}

\subsection{Self-consistent formation of an inclined current sheet\label{S:disc.current.sheet}}

One of the main goals of this study is to answer the question of whether or not footpoint motions at the solar surface can create a current sheet higher up in the atmosphere in which reconnection would lead to a bi-directional outflow like in an explosive event or UV burst.
The short answer is yes.

There have been other models of reconnection events in the upper atmosphere for explosive events or UV bursts.
But these are idealized in so far as they assume that oppositely directed magnetic fields already exist in the initial configuration of the numerical experiment, for example in the form of a Harris-type current sheet (see \sect{S:intro}).

The initial condition in our model with the X-type neutral point would be stable if we did not drive the system. Only after the footpoint driving starts, is the X-point stretched into a current sheet.
Thus the current sheet in our system forms self-consistently, which is a major difference as compared to earlier models for explosive events (as outlined in \sect{S:intro}).
Essentially, this process first leads to a (slow) Sweet-Parker-type reconnection, after which the plasmoid instability begins to take effect and greatly enhances the efficiency of the energy conversion process.

Considering the geometry of the magnetic setup (\figs{F:setup} and \fig{F:evolution}b-d), it is natural that the resulting current sheet will be inclined to the vertical.
Earlier models with a gravitational stratification were experimenting with cases where the current sheet was horizontal or vertical \cite[e.g.,][]{2016ApJ...832..195N}, but they simply assumed an orientation.
The magnetic setup we use in our model matches not only the observations by \cite{2017A&A...605A..49C} but also the more general observation that explosive events typically happen at locations where opposite polarities come into close contact and cancel \cite[][]{1998ApJ...497L.109C}.
Therefore, we can assume that our finding of an inclined current sheet in explosive events is quite general.
Consequently, the (bi-directional) outflow from the reconnection site will also be inclined, as has been inferred by the spectroscopic observations of \cite{1997Natur.386..811I}.
Such inclined outflows have also been suggested for UV bursts \cite[see Sect.\,S3 and Fig.\,S7 in the supplemental material of][]{2014Sci...346C.315P}.

We also see that the orientation (or inclination) of the current sheet in our model is consistent with the observations of \cite{2017A&A...605A..49C}.
In our model, the upward-directed part of the outflow is pointing away from the main polarity that the small opposite polarity is moving into.   
In the observations of  \cite{2017A&A...605A..49C}, the blueshifted, i.e., upflowing, part of the UV bursts is located on the side of the UV bursts pointing away from the pore, that is, the main polarity.
This is yet another nice match between model and observations, even though the analysis of more observational cases is needed to finally confirm this.

\subsection{Ultraviolet bursts in a low-$\beta$ environment: temporal evolution and plasmoids\label{S:disc.beta}}

The reconnection in our model is driven by the motion at the bottom boundary (i.e., \ the solar surface) moving the small opposite-polarity magnetic patch into the main polarity.
As such, the timescale of the driving determines the overall evolution of the system.
Still, independent of the driving, the dynamics in the reconnection region lead to a temporal variability on shorter timescales (${\approx}1$\,min) controlled by plasmoid formation.
This is consistent with observations of UV bursts.

We find that the conversion of energy in the reconnection region lasts for about 1000\,s (\fig{F:evolution.beta}, \sect{S:res.evolution}).
Once the driving stretched the current sheet sufficiently, the plasmoid-mediated reconnection begins and lasts as long as we drive (\sect{S:curr.sheet.plasmoids}).
While the duration of the reconnection event follows from our choice of the model parameters, our choice for the  duration of the driving is not completely free.
Instead, the driving time we choose is governed by observations, or more precisely by the (horizontal) magneto-convective motions on the surface.
In the UV burst discussed by \cite{2017A&A...605A..49C} the timescale of driving the small magnetic patch into the pore is about 10 to 20\,min; hence our choice of about 20\,min for the driving time. In a more general context, this timescale corresponds to a motion with about 1\,km\,s$^{-1}$ over a distance of about 1\,Mm.
These velocities and length scales are typical for granular magneto-convection near the surface.
Thus we can also consider our assumption for the driving time to be typical in cases where magnetic elements are pushed into (opposite) magnetic patches, for example\ with explosive events in the network \cite[][]{1998ApJ...497L.109C}.

During the time of the reconnection we see a clear  bi-directional outflow from the reconnection region.
This is consistent with observations of enhanced wings or a bi-modal shape of line profiles that is seen in explosive events \cite[e.g.,][]{1989SoPh..123...41D} or in part of the UV bursts \cite[e.g.,][]{2014Sci...346C.315P}.
At the same time, our model also shows plasmoids forming along the current sheets that can explain the enhanced emission observed in the line core, similar to the model of \cite{2015ApJ...813...86I}.

\begin{figure*}
\centerline{\includegraphics[width=150mm]{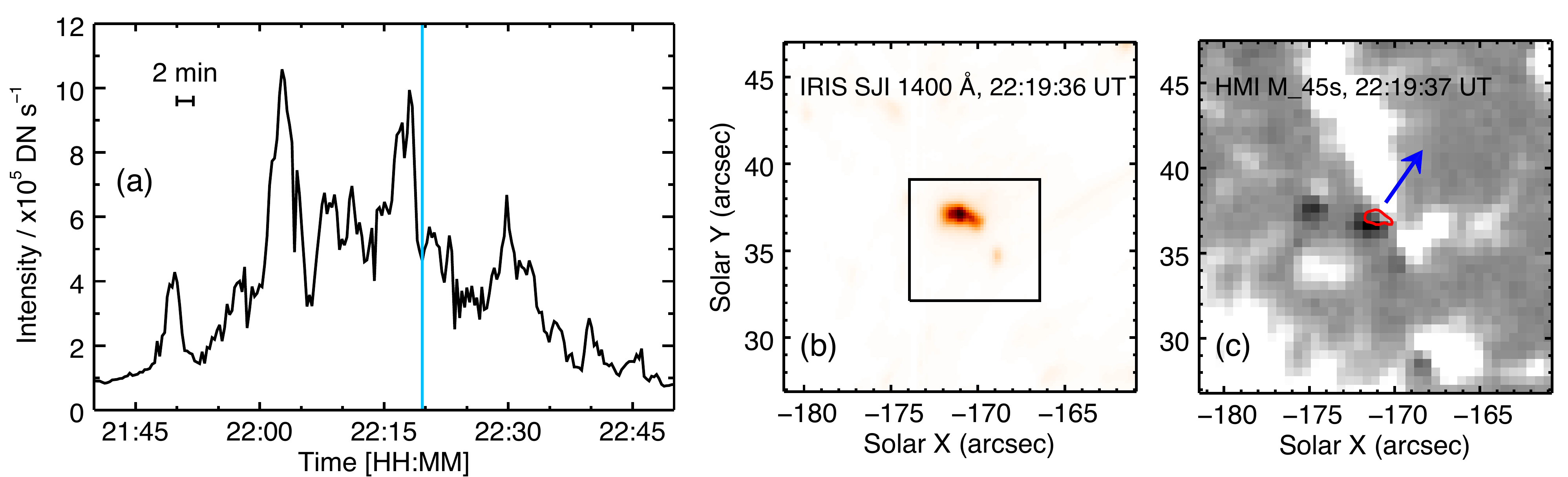}}
\caption{Temporal evolution of a solar UV burst observed by IRIS on 2013 October 22.
The temporal variation of the intensity in the IRIS 1400\,{\AA} channel essentially showing the light curve in the \ion{Si}{IV} line is plotted in panel (a).
Panel (b) shows the immediate surroundings of the UV burst as seen in the 1400\,{\AA} channel (reverse linear intensity scaling) at the time indicated by the vertical blue line in panel (a).
The box in panel (b) surrounding the UV burst outlines the region where we integrated the emission to derive the light curve in panel (a).
The line-of-sight component of the magnetic field based on HMI is shown in panel (c) with a linear scaling of the magnetic field in the range $\pm$200\,G. 
The location of the UV burst is over-plotted with a red contour of the 1400\,{\AA} channel intensity.
This only appears between a small negative (dark) magnetic polarity and a major opposite (white) polarity.
The blue arrow indicates the direction of motion of the minor negative-polarity feature that causes the UV burst.
See \sect{S:disc.beta} for details.
\label{F:obs.burst}}
\end{figure*}

What we find in our study is that the plasmoids are key to understanding the substructure seen in the temporal evolution of UV bursts, that is,\ their light curve. In \fig{F:obs.burst} we provide an example for the intensity variability during a UV burst. 
This shows that the event lasts for about one half to three quarters of an hour with a large number of local peaks in intensity, each lasting one or a few minutes. Another example is shown in Fig.\,3 of \cite{2018SSRv..214..120Y} with substructures spaced by some 30\,s.
Yet another case is discussed in  \cite{2015ApJ...809...82G} who found a flickering with a period from 30 to 90\,s.
In our models we see a  similar  sub-structure on timescales of 1 to 2\,min in all quantities derived from the reconnection region (\figs{F:zoom}a, \ref{F:evolution.beta}, \ref{F:T.plasma.b}a), and this substructure can be clearly related to the presence of plasmoids (\sect{S:res.general}).
Assuming that the UV bursts are driven by reconnection, it stands to reason that the observed flickering in UV bursts is also caused by plasmoids, with the enhanced temperature and density in the plasmoids leading to the enhancements in intensity (see also \sect{S:disc.int}).

The plasmoids move along the current sheet reaching (almost) the local Alfv\'en speed.
In our setup the Alfv\'en crossing time along the current sheet is of the order of 10 to 20\,s.
However, this  can only be a lower limit for the crossing of the plasmoids.
When formed, the plasmoid first has to accelerate and only when almost at the end of the current sheet hitting the ambient media does it reach the Alfv\'en speed.
Typically the time it takes the plasmoid to cross from the middle of the current sheet to its end is about 1 min.
The density and magnetic field in our model are motivated by a UV burst observation, and hence we expect that the (local) Alfv\'en speed in a UV burst is comparable to what we find in our model.
We therefore conclude that the plasmoids are indeed responsible for the 30 to 90\,s flickering seen in UV bursts.

In our model, at any given time we see one or two (large) plasmoids in the current sheet (e.g.,\ \fig{F:zoom}b).
This would also hold if we were to increase the spatial resolution.
For higher resolution, one expects that there are still large plasmoids, typically one in each current sheet at any given time, but these are now accompanied by smaller ones in a self-similar fashion \cite[][]{2012PhRvL.109z5002H}.
Essentially, at (much) higher resolution we would still see the clear peaks in all quantities of about one minute in length due to the (large) plasmoids, similar to the observed sub-structure in UV bursts. But now a noise-like pattern on much shorter timescales would be superimposed by plasmoids over a wide range of smaller scales.

\subsection{Plasmoids triggering fast-mode waves\label{S:disc.wave}}

The plasmoids that appear in a quasi-periodic pattern also launch fast
magneto-acoustic waves.
A more thorough analysis of this phenomenon would deviate from the main goal of this study.
Therefore, here we only refer to the movie attached to \fig{F:evolution} that shows the field lines shaking back and forth in a fashion expected for a fast-mode wave propagating away from the reconnection site.
Recently, such a phenomenon was observed
in the corona above the limb by \cite{2018ApJ...868L..33L}, albeit at larger scales.
These latter authors found a fast-mode wave with a period of about 4 min lasting for
many hours.
In their observation they find evidence for reconnection lasting
all that time, so we can speculate that plasmoids forming in the process
could give rise to the excitation of  fast-mode waves over such a long
time in a coherent fashion.

\subsection{Intensity from plasmoids compared to UV bursts\label{S:disc.int}}

The reconnection events we find in our model are energetic enough to power a typical UV burst.
At least, the intensity emerging from a plasmoid is consistent with the intensity observed in solar UV\ bursts.

In the plasmoids of our run with $\beta{=}0.007$ we typically find a peak temperature of about 50\,000\,K and densities of about $5{\times}10^{11}$\,cm$^{-3}$.
Assuming a line-of-sight length of 0.2\,Mm (representing the size of the plasmoid; cf.\ \fig{F:current.sheets}), we calculate an intensity radiating from the structure of about $10^{5}$\,erg\,cm$^{-2}$\,s$^{-1}$\,sr$^{-1}$.
For this estimate we use the Chianti database \cite[][]{1997A&AS..125..149D} and the procedure as described in \cite{2006ApJ...638.1086P}.
This fits well with the value of $5{\times}10^{5}$\,erg\,cm$^{-2}$\,s$^{-1}$\,sr$^{-1}$ for the UV burst observations quoted by \cite{2018SSRv..214..120Y} in their Sect.\,4.
We therefore conclude that our model can also give a quantitative explanation for the UV burst intensity.

\subsection{Why are explosive events restricted to temperatures below a few 10$^{\,5}$\,K?\label{S:disc.temp}}

As mentioned in the introduction, the question remains open as to why explosive events are observed only in spectral lines forming below about 0.4\,MK.
Here, our results on how the peak temperature in the reconnection region depends on plasma-$\beta$ are instructive (\sect{S:res.beta}, \fig{F:T.plasma.b}b) and can provide an answer.

Clearly, we find a monotonic increase of the peak temperature for smaller $\beta$.
In short, if plasma-$\beta$ is smaller in the reconnection region, there are less particles to be heated and thus the temperature can reach higher values.
While the upper solar atmosphere is a low-$\beta$ plasma, $\beta$ cannot be arbitrarily small and thus there is an upper limit for the temperature to be reached (realistically) during an explosive event.
Of course, this is particular for this model, and in a different setup, for example with continued flux emergence and higher driving speeds supported for longer times, one might reach coronal temperatures in the event \cite[][]{2018ApJ...864..165W}.

From a self-consistent 3D MHD model of the upper solar atmosphere one can derive plasma-$\beta$.
In their model, \cite{2006ApJ...638.1086P} showed that the smallest values of $\beta$ are found in the (low) transition region at temperatures from $10^4$\,K to  $10^5$\,K.
Typically, there  $\beta$ is larger than $10^{-3}$, and essentially never below $10^{-4}$ \cite[Fig.\,12c of][]{2006ApJ...638.1086P}.
Therefore, we can consider $10^{-4}$ as a lower limit for plasma-$\beta$.
In particular, this lower limit is also applied in the source region of typical transition region lines such as \ion{C}{IV} \cite[Fig.\,12a of][]{2006ApJ...638.1086P} and \ion{Si}{IV}.

We can now extrapolate the peak temperatures down to the smallest plasma-$\beta$ values to be expected and use that temperature as an upper limit for the temperatures to be expected in explosive events and UV bursts.
The lowest plasma-$\beta$ case we have in our numerical experiments is about 0.007.
Extrapolating using the power law based on \fig{F:T.plasma.b}b, we find values of the peak temperature of just below 0.1\,MK for $\beta{=}10^{-3}$ and  0.2\,MK for $\beta{=}10^{-4}$.
Consequently, we would not expect temperatures in UV bursts and explosive events to reach values much higher than 0.2 MK.
On the real Sun this upper limit might be slightly higher, because we most likely underestimate the temperatures in the reconnection region as discussed in \sect{S:model.eqs} following \eqn{E:energy}.
Therefore we consider our finding of a maximum reachable temperature in UV bursts and explosive events to be consistent with the observations.

Of course, this result applies only to situations for a driving as assumed in our model, that is, if one small patch of one magnetic polarity is moving into and canceling a larger patch of opposite polarity.
Under other circumstances the driving of the magnetic field might be located closer to the reconnection location, as expected for Ellerman bombs or some UV bursts \cite[see cartoons in][]{2002ApJ...575..506G,2014Sci...346C.315P}.
A model might therefore produce much higher temperatures. Still,  such hot plasma, even if reaching coronal temperatures, might not show up in extreme UV emission (e.g., in the widely used 171\,{\AA} band showing mostly coronal plasma in \ion{Fe}{iv}).
This is because of absorption by overlying cooler material in the Lyman-continua of hydrogen and helium, \cite[cf.\ suppl.\ material SM2 of][]{2014Sci...346C.315P}.\

\subsection{Relation to Ellerman bombs and high-$\beta$ reconnection\label{S:disc.EB}}

The particular configuration we use here cannot explain more violent bursts
at deeper atmospheric layers, that is, events that originate where plasma-$\beta$ is close to unity or even larger.
For the particular driving mechanism we employ here, we do not find any significant increase of the kinetic energy or of the temperature during the reconnection event for values of $\beta$  larger than about 0.5 (cf.\ \figs{F:evolution.beta} and \ref{F:T.plasma.b}).
When driving the magnetic configuration from the footpoints, the inertia of the plasma can hinder the propagation of the changes of the magnetic field into the upper atmosphere (\sect{S:res.kinetic}).
Consequently, there is only a small amount of energy conversion and hence no observational consequences are expected if plasma-$\beta$ is close to or above unity.

Of course, reconnection can happen also in a high-$\beta$ plasma \cite[e.g.,][]{2010PhPl...17f2104H}.
However, in this situation the reconnection has to be driven by a flow of plasma converging in the reconnection region. This is exactly the scenario that recent 3D models \cite[][]{2017ApJ...839...22H,2017A&A...601A.122D} suggest for Ellerman bombs originating in the photosphere, that is, in a region where $\beta$ is above unity on average\textbf{}.
Here the converging horizontal flows arising from the granular convective motions push together opposite magnetic field which then drives magnetic reconnection \cite[][]{2017A&A...601A.122D}.

In a configuration like the one we use here to study explosive events and UV bursts we cannot explain Ellerman bombs.
This is because the (horizontal) motions in the high-$\beta$ regions that push together oppositely directed magnetic field are not present.
On the other hand, in a scenario in which an Ellerman bomb is initiated, one might well expect this to also induce effects at higher temperatures resembling UV bursts; this has indeed been suggested recently based on observations by \cite{2019ApJ...875L..30C}. Therefore, one might speculate that a higher fraction of Ellerman bombs is accompanied by UV bursts, rather than the other way round.
However, more modeling and observational efforts  are needed to conclude on the processes connecting these transient events.

\subsection{Applicability of the model to UV bursts\label{S:limitation}}

One major limitation of the model is the incompleteness of the physical processes in the chromosphere. 
Observations indicate that UV bursts are launched somewhere between the temperature minimum and the middle chromosphere \cite[][]{2014Sci...346C.315P,2016ApJ...824...96T,2018ApJ...854..174T,2017A&A...605A..49C}.
In response to the reconnection, the local plasma gets heated and radiates in emission lines such as \ion{Si}{iv}.
Therefore, during the onset of the reconnection, we have to expect the plasma to be only partially ionized.
This would require consideration of multi-fluid effects, as was the case in recent numerical reconnection experiments \cite[][]{2018ApJ...852...95N,2018PhPl...25d2903N,2018ApJ...868..144N}.
These models showed that nonequilibrium (partial) ionization alters the radiative cooling and thus can  significantly affect the temperature evolution in the reconnection region.
In consequence, the temperature increase would be lower than what we find in our numerical experiments that do not consider multi-fluid effects.

If we included multi-fluid effects and set off the reconnection near the temperature minimum, i.e., below 5000\,K, the peak temperature would be lower than what we find in our study.
However, not all transients seen in the extreme UV will originate at the bottom of the chromosphere.
For example, explosive events seen in the quiet Sun are generally thought to be reconnection events starting at lower densities (i.e., higher temperatures) than what is found in the chromosphere.
For those events, as we discuss in \sect{S:disc.temp}, our setup with an initial temperature of $10^4$\,K, that is, well above the temperature minimum, might provide a more realistic temperature estimate.

\section{Conclusions\label{S:conclusions}}

In our study we investigate  a reconnection model around  an X-type neutral point suitable to understand the dynamics in an explosive event or UV burst.
The magnetic setup is motivated by observations showing that such events usually take place in situations where a (small)  patch of one magnetic polarity is moving into a (main) patch of opposite polarity.
In particular, we select our simulation parameters based on the observations of \cite{2017A&A...605A..49C}.

The driving stretches the X-point into a current sheet and quickly the plasmoid instability takes effect.
The enhancement of for example kinetic energy and temperature in the reconnection region is very closely related to the presence of plasmoids. 
In contrast to earlier studies, in our model the current sheet forms self-consistently in response to the driving of the (small) magnetic patch at the bottom boundary of our computational box, that is,\ close to the solar surface (\sect{S:disc.current.sheet}).

The enhancement of the energy deposition essentially lasts as long as we drive the system. However, we find a fine structure in the temporal evolution with short busts lasting for around one minute. 
This intermittency is governed by the plasmoids, and the timescale corresponds to the Alfv\'en crossing time along the current sheet.
This corresponds very well to the observed sub-structure (or flickering) of the light curve observed in UV bursts (\sect{S:disc.beta}).
Estimating the radiative losses from our reconnection experiments, we find that these roughly match the observations of UV bursts (\sect{S:disc.int}).

One main goal of our study was to investigate the effect of plasma-$\beta$ on the reconnection.
For high values of $\beta$, that is,\ if the thermal energy dominates, we find that this reconnection process driven from the surface is not efficient.
In this case, the inertia of the plasma can hinder the changes of the magnetic field in reaching the X-type neutral point. 
For low $\beta$, that is,\ if the magnetic field dominates, the driving has an increasing effect and hence the resulting temperatures and the velocities found in the reconnection region increase with decreasing $\beta$. 
However, we cannot expect this process to reach arbitrarily high temperatures because generally, plasma-$\beta$ will never drop below $10^{-4}$ in the solar atmosphere. 
Therefore, practically, we find that the temperatures in the reconnection region should not reach values significantly above 10$^5$\,K. 
This is in accordance with observations that show that explosive events are essentially limited to a narrow temperature range of about 10$^5$\,K.
Still, under certain circumstances in UV bursts, with different driving from that applied here, one might expect coronal temperatures (\sect{S:disc.temp}).

Our reconnection models provide new insight into the physics of explosive events and UV bursts. 
These can be driven by motions of magnetic patches at the solar surface, self-consistently resulting in current sheets and plasmoids. 
Using this approach we reproduce key features such as duration, temporal sub-structure of the light curve (flickering), and the preferential temperatures of explosive events around 10$^5$\,K.

\begin{acknowledgements}
We gratefully acknowledge the constructive comments from the referee.
L.P.C. received funding from the European Union's Horizon 2020 research and innovation programme under the Marie Sk{\l}odowska-Curie grant agreement No. 707837.
This work was supported by the Max Planck/Princeton Center for Plasma Physics sponsoring trips of H.P., L.P.C, and Y.-M.H.
The simulations were performed using facilities of the National Energy Research Scientific Computing Center.
H.P.\ acknowledges the hospitality at the Princeton Plasma Physics Laboratory during his stay.
P.R.Y.\ acknowledges funding from NASA grant NNX15AF48G, and thanks the Max Planck Institute in G\"ottingen for kind hospitality during visits in 2015 and 2016. Y.-M.H.\ acknowledges funding from NSF grant AGS-1460169, DOE grant DE-SC0016470, and NASA grant 80NSSC18K1285, and thanks the Max Planck Institute in G\"ottingen for kind hospitality during visits in 2016 and 2017.
This research has made use of NASA's Astrophysics Data System.
\end{acknowledgements}

\end{document}